\documentclass[]{spie}  
\usepackage[]{graphicx}
\usepackage{amsmath}
\usepackage{enumerate}
\usepackage{color}
\usepackage{cite}
\usepackage{mathptmx}      
\usepackage{bm}
\usepackage{lineno}

\title{Quantum theory as a description of robust experiments:
Application to Stern-Gerlach and Einstein-Podolsky-Rosen-Bohm experiments}

\author{H. De Raedt\supit{a}, M.I. Katsnelson\supit{b}, H.C. Donker\supit{b}, and K. Michielsen\supit{c,}\supit{d}
\skiplinehalf
\supit{a}
Zernike Institute for Advanced Materials, University of Groningen, \\
NL-9747 AG Groningen, The Netherlands
\\
\supit{b}
Institute of Molecules and Materials, Radboud University of Nijmegen,\\
NL-6525 ED Nijmegen, The Netherlands
\\
\supit{c}
Institute for Advanced Simulation, J\"ulich Supercomputing Centre,
Forschungszentrum J\"ulich, D-52425 J\"ulich, Germany
\\
\supit{d}
RWTH Aachen University, D-52056 Aachen, Germany
}

\authorinfo{
Further author information: (Send correspondence to H. De Raedt)\\
H. De Raedt : E-mail: h.a.de.raedt@rug.nl \\
M.I. Katsnelson: E-mail: M.Katsnelson@science.ru.nl \\
H.C. Donker: E-mail: H.Donker@science.ru.nl \\
K. Michielsen: E-mail: k.michielsen@fz-juelich.de}

\DeclareRobustCommand\openone{\leavevmode\hbox{\small1\normalsize\kern-.33em1}}
\newcommand{\onlinecite}{\cite}
\newcommand{\url}[1]{{$\mathrm{#1}$}}

\begin{document}
  \maketitle

\begin{abstract}
We propose and develop the thesis that the
quantum theoretical description of experiments emerges from the desire
to organize experimental data such that
the description of the system under scrutiny and the one used to acquire the data
are separated   as much as possible.
Application to the Stern-Gerlach and Einstein-Podolsky-Rosen-Bohm
experiments are shown to support this thesis.
General principles of logical inference
which have been shown to lead to the Schr\"odinger and Pauli equation
and the probabilistic descriptions of the Stern-Gerlach and Einstein-Podolsky-Rosen-Bohm experiments,
are used to demonstrate that the condition for the separation
procedure to yield the quantum theoretical description
is intimately related to the assumptions that
the observed events are independent and that
the data generated by these experiments is robust with
respect to small changes of the conditions under which the experiment is carried out.
\end{abstract}

\keywords{logical inference, quantum theory, inductive logic, probability theory}

\section{Introduction}\label{sec1}

As quantum theory has proven remarkably powerful to describe many very different experiments in
(sub)-atomic, molecular and condensed matter physics, quantum optics and so on,
it is of interest to search for explanations that go beyond ``that is because of the way it is''.
Recently, several papers~\cite{RAED13b,RAED14b,RAED15b} suggest that such an explanation can be given.
Some of the most basic elements of quantum theory, such as the Schr\"odinger equation, have been derived~\cite{RAED13b,RAED14b,RAED15b}
by taking the mathematical framework of logical inference (LI)~\cite{COX46,COX61,TRIB69,SMIT89,JAYN03}
as a basis for establishing a bridge between the data gathered through experiments and their description in terms of (mathematical) concepts.
These papers are not concerned with the various interpretations of quantum theory~\cite{KHRE09,HOME97,BALL01,GRIF02}
but employ a mathematically precise formalism that expresses what most people would consider to be rational reasoning.
The key concept is the notion of the plausibility that a proposition is true~\cite{COX46,COX61,TRIB69,SMIT89,JAYN03}.

The basic premise of the LI approach~\cite{RAED13b,RAED14b,RAED15b} is that
scientific theories are created through cognitive processes in the human brain,
based on discrete events which are observed in every-day life and in laboratory experiments
combined with the logical and/or cause-and-effect relations between those events that we, humans, uncover.
Because of this premise, the LI approach does not depend on assumptions such as that
the observed events are signatures of an underlying objective reality
-- which may or may not be mathematical in nature,
that all things physical are information-theoretic in origin,
that the universe is participatory etc.

The rules of LI are not bound by ``laws of physics''.
LI applies to situations where there may or may not be causal relations between the events~\cite{TRIB69,JAYN03}.
Although extracting cause-and-effect relationships from empirical evidence
is a highly non-trivial problem, rational reasoning about these relations
should comply with the rules of LI. However, in general, the latter cannot
be used to establish cause-and-effect relationships themselves~\cite{JAYN03,PEAR09,PLOT11b}.
A derivation of a quantum theoretical description from LI principles
does not prohibit the construction of cause-and-effect mechanisms that create the {\it impression}
that these mechanisms produce data that can be described by quantum theory~\cite{RAED05b}.
In fact, there is a substantial body of work demonstrating that it is indeed possible to build simulation models
which reproduce, on an event-by-event basis, the results of interference/entanglement/uncertainty
experiments with single photons/neutrons~\cite{MICH11a,RAED12a,RAED12b,RAED14a,MICH14a}.

In the LI approach it is not necessary to assume that the observed events are ``generated'' according to some (quantum) laws.
These laws naturally emerge from inference based on available data.
There is no need to postulate ``wavefunctions'', ``wavefunction collapse'',
``observables'', ``quantization rules'', ``Born's rule'', ``wave-particle duality'' etc.
and there is no ``quantum measurement problem''~\cite{NIEU13}.
For a particle in a (electromagnetic) potential, the solution of the inference problem
takes the form of a set of non-linear equations and the
Schr\"odinger and Pauli equation appear as a result of transforming this set
into a set of linear equations which are much more convenient for further analysis~\cite{RAED13b,RAED14b,RAED15b}.

The mathematical framework of LI~\cite{COX46,COX61,TRIB69,SMIT89,JAYN03} applied
to basic problems such as the idealized Stern-Gerlach (SG) and Einstein-Podolsky-Rosen-Bohm (EPRB) experiments
also yield the probability distributions that we know from quantum physics~\cite{RAED14b}.
However, unlike in the case of the particle in a potential where
the Schr\"odinger and Pauli equation appear as a result of transforming a set of non-linear equations
into a set of linear equations, for the SG and EPRB experiment {Ref.~\onlinecite{RAED14b}}
did not explicitly derive the wavefunction (density matrix) description but was content by showing that
the robust solution of the inference problem yields the expressions of the
probabilities that are known from quantum theory~\cite{RAED14b}.
In this paper, we fill this void by demonstrating, using
the SG and EPRB experiment as concrete examples, how the mathematical structure of quantum theory
naturally emerges from the desire to organize experimental data such that it separates    as
much as possible, the description of the system under scrutiny
from the description of the probe used to acquire the data.

The idea that the mathematical structure of quantum theory has application to cognitive science, psychology,
genetics, economics, finances, and game theory is discussed in great depth by Khrennikov~\cite{KHRE10}.
Building on the ``wavefunction postulate'', Khrennikov shows how to construct contextual probabilistic
models of natural, biological, psychological, social, economical or financial phenomena~\cite{KHRE10}.
In particular, Born's rule appears by representing probabilistic data in terms of complex probability amplitudes~\cite{KHRE10}.
The route taken in the present paper is complementary and more along the line of thinking
explored in {Ref.~\onlinecite{RALS13b}}.

A general, characteristic feature of scientific reasoning is that
it strives to reduce the complexity of the description of the whole
by separating the description of data into several independent parts.
The separation of variables when solving differential equations,
the Fourier analysis of a signal, a principal component analysis of a correlation matrix of data,
the normal mode description of the vibrations in a harmonic solid,
are just a few of the vast set of instances where this division is used to great advantage.
In the present paper, we start by considering different ways of organizing observed data
and scrutinize the conditions for which the data can be represented such that
the description of the various components of the experimental setup can be separated  as much as possible.
We show that the wavefunction description appears as the result of such a separation  procedure.
It automatically follows that the wavefunction (or density matrix) description
is less general than the one in terms of conditional probabilities in the sense that
the former can only describe situations in which the separation   procedure can actually be carried out.

The paper is structured as follows.
Section~\ref{sec2} starts with the analysis of data produced by an idealized SG experiment.
We show how compression of data and subsequent reorganization offers
the possibility to separate the description of the source and the magnet
in this particular case, leading to the quantum theoretical description in terms
of a wavefunction of a spin-1/2 system.
Then, we briefly recapitulate how the LI approach directly leads
to the probabilities for observing the events and show how
the results of quantum theory emerge without invoking any of its postulates.
In Section~\ref{sec3}, we apply the same logic to the EPRB thought experiment
and demonstrate how the singlet state emerges by rearranging the data
such that separation   becomes possible.
Although this derivation does not add anything new to the ongoing discussions about locality, realism, etc.
in relation to the violation of Bell-like inequalities~\cite{PENA72,FINE74,BROD93,KUPC86,JAYN89,SICA99,
HESS01a,HESS05,KRAC05,SANT05,NIEU09,KARL09,NIEU11,RAED11a,HESS15},
it does de-mystify the ``origin'' and ``nature'' of quantum entanglement.

\section{Idealized Stern-Gerlach experiment}\label{sec2}

\begin{figure}
\includegraphics[width=\hsize]{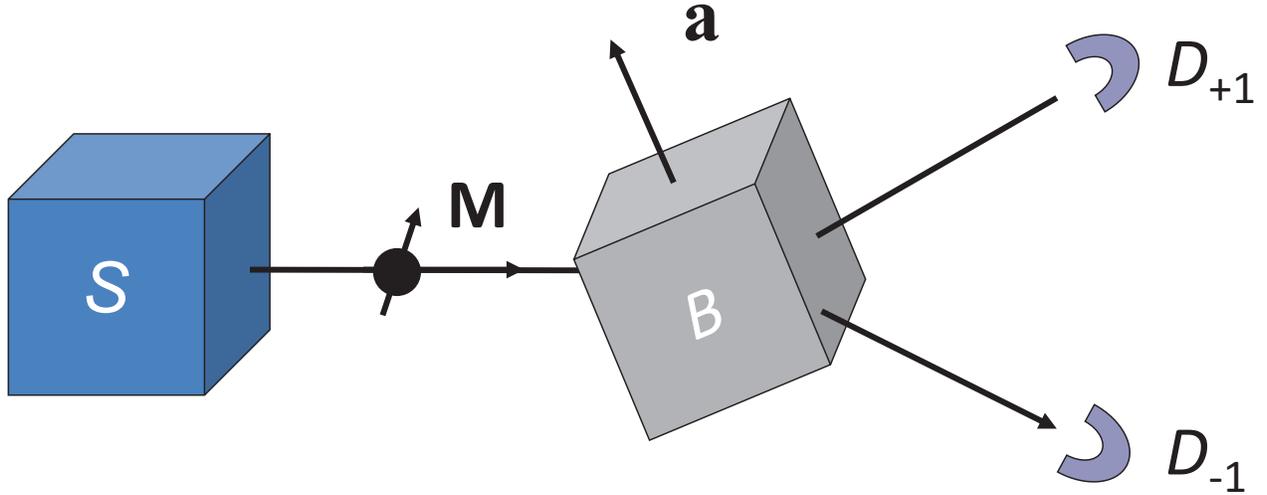}
\caption{(Color online)
Diagram of the SG experiment.
The source $S$, activated at times labeled by $i=1,2,\ldots,N$,
sends a particle carrying a magnetic moment $\mathbf{M}$
to the magnet $B$ with its (inhomogeneous) magnetic field along the direction $\mathbf{a}$.
Depending on the relative directions of $\mathbf{a}$ and $\mathbf{M}$,
the particle is detected with 100\% certainty by either
$D_{+1}$ or $D_{-1}$.
}
\label{SGthought}
\end{figure}

In the idealized SG experiment, a schematic of which is shown in Fig.~\ref{SGthought},
there are two different outcomes which we label by the variable $x$ taking
the values $+1$ or $-1$.
We imagine that the experiment is repeated $N$ times, yielding the data set
\begin{eqnarray}
{\cal D} &=& \big\{x_1,\ldots,x_N|x_i=\pm1\;,i=1,\ldots,N\big\}
.
\label{SG0}
\end{eqnarray}
Assuming that the observed $x_i$ are independent events (uncorrelated in time),
the data is completely characterized by the counts
$N(+1|\mathbf{a},\mathbf{M},Z)$ of outcomes with
$x=+1$ and $N(-1|\mathbf{a},\mathbf{M},Z)$ of outcomes with $x=-1$
where $N=N(+1|\mathbf{a},\mathbf{M},Z)+N(-1|\mathbf{a},\mathbf{M},Z)$.
Here and in the following, we use the notation ``$|a,b,\ldots)$'' to indicate that the data
is collected under the ${\it fixed}$ conditions $a,b,\ldots$.
In the case at hand, $\mathbf{a}$, $\mathbf{M}$, and $Z$ represent, respectively, a unit vector
specifying the direction of the SG magnet,
a unit vector representing the direction of the magnetic moments of the incoming particles,
and all other fixed conditions
which are important to the actual experiment but are not of immediate interest.
Assuming independent $x_i$'s, the average reads
\begin{eqnarray}
\langle x \rangle &=& \sum_{x=\pm1} x f(x|\mathbf{a},\mathbf{M},Z)
\quad,\quad
f(x|\mathbf{a},\mathbf{M},Z)\equiv N(x|\mathbf{a},\mathbf{M},Z)/N
,
\label{SG1}
\end{eqnarray}
where $f(x|\mathbf{a},\mathbf{M},Z)$
denotes the (empirical) frequency of observing the event $x$ and
completely characterizes the outcomes of an idealized SG experiment of $N$ events.
As $f(x|\mathbf{a},\mathbf{M},Z)$ is a function of the two-valued variable $x$,
it can be written as
\begin{eqnarray}
f(x|\mathbf{a},\mathbf{M},Z)&=&\frac{1+xE(\mathbf{a},\mathbf{M},Z)}{2}
,
\label{INT1}
\end{eqnarray}
where $E(\mathbf{a},\mathbf{M},Z)=\langle x \rangle$.
Assuming that the observed counts do not depend on the orientation
of the chosen reference frame,
$E(\mathbf{a},\mathbf{M},Z)$ can only depend on $\mathbf{a}\cdot\mathbf{M}$
(by construction $\vert\mathbf{a}\vert=1$ and $\vert\mathbf{M}\vert=1$).
Hence, we must have $f(x|\mathbf{a},\mathbf{M},Z)=f(x|\mathbf{a}\cdot\mathbf{M},Z)$.

\subsection{Separation  procedure}\label{sec2a}

It is obvious from Eq.~(\ref{SG1}) that the description in terms of $x$ and $f(x|\mathbf{a},\mathbf{M},Z)$
does not readily allow for the separation   that we envisage.
Therefore, it is necessary to consider rewritings of Eq.~(\ref{SG1}) that lend themselves to such a separation.

Let us organize data of observations and frequencies in vectors $\mathbf{x}=(+1,-1)^T$
and $\mathbf{f}=(f(+1|\mathbf{a},\mathbf{M},Z),f(-1|\mathbf{a},\mathbf{M},Z))^T$, respectively.
Then, we have
\begin{eqnarray}
\langle x \rangle &=& \mathbf{x}^T\cdot\mathbf{f}= \mathbf{Tr\;} \mathbf{x}^T\mathbf{f}= \mathbf{Tr\;}\mathbf{f} \mathbf{x}^T
,
\label{P2D0a}
\end{eqnarray}
where $\mathbf{f} \mathbf{x}^T$ is a $2\times2$ matrix
and $\mathbf{Tr}\; A $ denotes the trace of the matrix $A$.

The key to the separation   procedure is to note that {\bf any} rewriting of $\mathbf{x}$
and $\mathbf{f}$ in terms of vectors, matrices, \ldots,
$\widetilde\mathbf{x}$ and $\widetilde\mathbf{f}$ such that
$\mathbf{Tr\;}\widetilde\mathbf{f}=1$ and
$\mathbf{Tr\;}\widetilde\mathbf{f}\widetilde\mathbf{x}=\langle x \rangle$ hold may be useful.
With this in mind, we consider the rearrangement of the data
into $2\times2$ (diagonal, hermitian) matrices $X$ and $F$ with elements
$X(x,x')=x\delta_{x,x'}$ and
$F(x,x')=f(x|\mathbf{a},\mathbf{M},Z)\delta_{x,x'}$, respectively.
Then, Eq.~(\ref{P2D0a}) can be written as
\begin{eqnarray}
\langle x \rangle &=& \mathbf{Tr}\; FX = \mathbf{Tr}\; \widehat\rho \widehat X
,
\label{P2D0}
\end{eqnarray}
where $\widehat\rho$ and $\widehat X$ can be any pair of $2\times2$ matrices
that satisfies Eq.~(\ref{P2D0}).
As Eq.~(\ref{P2D0}) is just a formal rewriting of Eq.~(\ref{P2D0a})
it can, by itself, not bring anything new.
Clearly, in the original representation of the data Eq.~(\ref{SG1}),
separation   is impossible (as it is implicitly assumed that $f(x|\mathbf{a},\mathbf{M},Z)$
does indeed depend on $\mathbf{a}$, $\mathbf{M}$, and $Z$).
However, the flexibility offered by representation Eq.~(\ref{P2D0})
allows us to perform the separation, as we now show.

From linear algebra, we know that any hermitian $2\times2$ matrix
can be written as a linear combination of four hermitian $2\times2$ matrices.
These four matrices may be chosen such that they are orthonormal
with respect to the inner product defined by $(A,B)=\mathbf{Tr}\; A^\dagger B$.
Using the Pauli-spin matrices $\bm{\sigma}=(\sigma^x,\sigma^y,\sigma^z)$
and the unit matrix $\openone$ as the orthonormal basis set for the vector space of
$2\times2$ matrices, we may write (without loss of generality)
\begin{equation}
\widehat\rho=\frac{\rho_0\openone+\bm\rho\cdot\bm{\sigma}}{2}
\quad,\quad
\widehat X=u_0\openone+\mathbf{u}\cdot\bm{\sigma}
.
\label{P2D1}
\end{equation}
where $\rho_0$, $\bm\rho=(\rho_x,\rho_y,\rho_z)$, $u_0$, and $\mathbf{u}=(u_x,u_y,u_z)$
are real-valued objects.
Using $(\mathbf{a}\cdot\bm{\sigma})\;(\mathbf{b}\cdot\bm{\sigma})=
\mathbf{a}\cdot\mathbf{b}+i(\mathbf{a}\times\mathbf{b})\cdot\bm{\sigma}$
we find
\begin{equation}
2\widehat\rho\widehat X
=u_0\rho_0\openone+u_0\bm\rho\cdot\bm{\sigma}+\rho_0\mathbf{u}\cdot\bm{\sigma}
+\bm\rho\cdot\mathbf{u}\openone+i(\bm\rho\times\mathbf{u})\cdot\bm{\sigma}
.
\label{P2D3}
\end{equation}
Making use of $\mathbf{Tr}\; \openone=2$ and $\mathbf{Tr}\; \bm{\sigma}=0$ we find
\begin{eqnarray}
\mathbf{Tr}\;\widehat\rho =\rho_0
\quad,\quad 
\mathbf{Tr}\;\widehat\rho\widehat X = \rho_0u_0+\bm\rho\cdot\mathbf{u}
.
\label{P2D4}
\end{eqnarray}
Note that unlike $FX$, the matrix Eq.~(\ref{P2D3}) is not hermitian.
Obviously, useful alternative representations of the data should not change the description of the data,
that is we should impose the constraints $\mathbf{Tr}\; F= \mathbf{Tr}\; \widehat\rho=1$
and $\mathbf{Tr}\; F X= \mathbf{Tr}\; \widehat\rho \widehat X$.
Hence, from Eqs.~(\ref{P2D0}) and~(\ref{P2D4}) it follows that
\begin{eqnarray}
\langle x \rangle&=&u_0+\bm{\rho}\cdot\mathbf{u}
,
\label{P2D5}
\end{eqnarray}
suggesting that the desired separation  can be realized by requiring that
$u_0=u_0(\mathbf{a},Z)$,
$u_x=u_x(\mathbf{a},Z)$,
$u_y=u_y(\mathbf{a},Z)$,
$u_z=u_z(\mathbf{a},Z)$,
$\rho_x=\rho_x(\mathbf{M},Z)$,
$\rho_y=\rho_y(\mathbf{M},Z)$,
and
$\rho_z=\rho_z(\mathbf{M},Z)$ (recall that $Z$ is considered to represent all fixed conditions
which are important to the actual experiment but are not of immediate interest).
Assuming that the observed counts do not depend on the orientation of the reference frame (see earlier),
$\langle x \rangle$ is a function of $\mathbf{a}\cdot\mathbf{M}$ only.
This requirement enforces $\bm{\rho}=\mathbf{M}$, and $\mathbf{u}=\mathbf{a}$.
Hence, we have
\begin{eqnarray}
\langle x \rangle&=&u_0+\mathbf{M}\cdot\mathbf{a}
.
\label{P2D5a}
\end{eqnarray}
As $|\langle x \rangle|\le 1$ it follows from Eq.~(\ref{P2D5a}) that
$-1\le u_0+\mathbf{M}\cdot\mathbf{a}\le1$.
As the unit vector $\mathbf{a}$ can take any value on
the unit sphere, we find that $u_0\le 0 $ and $0\ge u_0$
for $\mathbf{a}=\mathbf{M}$ and $\mathbf{a}=-\mathbf{M}$, respectively,
hence $u_0=0$.
Note that we have tactically written $\bm{\rho}=\mathbf{M}$, and $\mathbf{u}=\mathbf{a}$
but it is clear that at this point, we could equally well
have made the choice $\bm{\rho}=\mathbf{a}$, and $\mathbf{u}=\mathbf{M}$.
However, the latter choice leads to inconsistencies when we consider
for instance an experiment in which we place several SG magnets
in succession or consider the EPRB experiment (see Section~\ref{sec3}).

We may therefore conclude that the desire
to represent the data Eq.~(\ref{SG0}) such that the description of the whole experiment
is separated into a description of the ``source'' ($\mathbf{M}$)
and a description of the ``measurement device'' ($\mathbf{a}$) leads to the unique representation
in terms of $2\times2$ matrices
\begin{equation}
\widehat\rho=(\openone+\mathbf{M}\cdot\bm{\sigma})/2
\quad,\quad
\widehat X=\mathbf{a}\cdot\bm{\sigma}
,
\label{P2D6}
\end{equation}
conditional on the assumptions that the individual outcomes of a SG experiment
are independent and that the frequency distribution of these outcomes
does not depend on the orientation of the reference frame.

From Eq.~(\ref{P2D6}) it follows immediately that $\widehat\rho^2=\widehat\rho$, that is $\widehat\rho$ is a projection.
This implies that we can write~\cite{BALL03}
\begin{equation}
\widehat\rho=|\Psi\rangle \langle\Psi|
\quad,\quad
|\Psi\rangle=a_\uparrow|\uparrow\rangle+a_\downarrow|\downarrow\rangle
.
\label{P2D6a}
\end{equation}
where the vector $|\Psi\rangle$ is expressed in the basis of the eigenstates ($|\uparrow\rangle$,$|\downarrow\rangle$)
of the $\sigma^z$ matrix.
Therefore, we may conclude that in the case of the SG experiment,
changing the representation of the
data in combination with the desire to separate   as much as possible the description of
the source and measurement devices automatically enforces the Hilbert space
structure that is a characteristic signature of quantum theory.

Starting from Eq.~(\ref{P2D6}) we can construct a mixed state~\cite{BALL03} by
multiplying Eq.~(\ref{P2D6}) with $w(\mathbf{M})d\mathbf{M}$ and integrating
over $\mathbf{M}$, $w(\mathbf{M})$ being the probability density of $\mathbf{M}$.
This changes Eq.~(\ref{P2D6}) into
\begin{equation}
\widehat\rho=(\openone+\langle\mathbf{M}\rangle\cdot\bm{\sigma})/2
\quad,\quad
\widehat X=\mathbf{a}\cdot\bm{\sigma}
,
\label{P2D6z}
\end{equation}
where $\langle\mathbf{M}\rangle=\int \mathbf{M}\;w(\mathbf{M})d\mathbf{M}$.
Then $\widehat\rho^2\not=\widehat\rho$ if $|\langle\mathbf{M}\rangle|<1$.
Therefore, under the conditions stated, the separation   procedure
yields the more general mixed-state description of a spin-1/2 object.
Applying the same reasoning to generalize the ``unseparated'' description,
we have
\begin{eqnarray}
\langle x\rangle
=\int\frac{1+xE(\mathbf{a}\cdot\mathbf{M},Z)}{2} w(\mathbf{M})d\mathbf{M}
=\frac{1+x\langle E(\mathbf{a}\cdot\mathbf{M},Z)\rangle}{2}
,
\label{P2D7}
\end{eqnarray}
where $\langle E(\mathbf{a}\cdot\mathbf{M},Z)\rangle=\int E(\mathbf{a}\cdot\mathbf{M},Z) w(\mathbf{M})d\mathbf{M}$.
In general,
$\langle E(\mathbf{a}\cdot\langle\mathbf{M}\rangle,Z)\rangle\not= E(\mathbf{a}\cdot\langle\mathbf{M}\rangle,Z)$.
Hence, even before we attempt to separate the descriptions,
it is clear that Eq.~(\ref{P2D7}) can describe experimental data that
cannot be represented by Eq.~(\ref{P2D6z}).
Moreover, there is nothing that forbids an experiment to yield for instance
$E(\mathbf{a}\cdot\langle\mathbf{M}\rangle),Z)=(\mathbf{a}\cdot\langle\mathbf{M}\rangle)^2$
(we certainly can generate such data using a digital computer, a physical
device on which we carry out experiments)
but the data produced by such an experiment cannot be represented by Eq.~(\ref{P2D6}).
In other words, the class of conceivable SG experiments is significantly larger
than the class of experiments that allows for the separation,
that is this class is (much) larger than the class of SG experiments describable by quantum theory.

Summarizing: we have derived the quantum theoretical description of the idealized SG experiment
from the desire to separate   the representation of the observed data
into a description of the object under study, its properties being
represented by the density matrix $\widehat\rho =\widehat\rho(\mathbf{M})$
and a description of the measurement device, its properties being represented by $\widehat X=\mathbf{a}\cdot\bm{\sigma}$.
Although the relation to quantum theory is obvious, our derivation does not invoke
any of the postulates of quantum theory but merely exploits
different representations of the recorded two-valued data, see Eq.~(\ref{SG0}).

The fact that the separation   procedures leads, in such simple manner,
to the quantum theoretical description Eq.~(\ref{P2D6z}) of the idealized SG experiment
provokes to question ``what is so special about the case in which the separation   procedure
can be carried out?'' The answer, as we show in the next subsection, is related
to the notion of robust experiments~\cite{RAED13b,RAED14b,RAED15b}.

\subsection{Logical-inference treatment}\label{sec2b}

For the reader's convenience, we briefly recapitulate the LI treatment
of the idealized SG experiment depicted in Fig.~\ref{SGthought}~\cite{RAED14b}.
The key concept in a LI treatment is the plausibility,
denoted by $P(A|B)$, an intermediate mental construct that
serves to carry out inductive logic, that is rational reasoning,
in a mathematically well-defined manner~\cite{TRIB69,JAYN03}.
In general, the plausibility $P(A|B)$ expresses the degree of believe of an individual that
proposition $A$ is true, given that proposition $B$ is true.
However, we explicitly exclude applications of this kind because they do not comply with our main goal,
namely to describe phenomena in a manner independent of individual subjective judgment~\cite{BOHR99}.
Therefore, we will refer to ``inference-probability'' or ``i-prob'' for short
to differentiate between the ``objective'' and ``subjective'' mode of application of LI.
A previous paper~\cite{RAED14b} gives detailed arguments for distinguishing between plausibility,
inference-probability, and Kolmogorov probability. For the present paper,
it suffices when the reader does not think of i-prob's as frequencies.

The first step in the LI treatment is to assign to an individual event,
an i-prob $P(x|\mathbf{a}\cdot\mathbf{M},Z)$ to observe an event $x=\pm1$
where $Z$ represents all the conditions under which the experiment is performed, with exception of
the directions $\mathbf{a}$ and $\mathbf{M}$ of the magnet and of the magnetic moment of the particle, respectively.
It is assumed that the conditions represented by $Z$ are fixed and identical for all experiments.
It is expedient to write $P(x|\mathbf{a}\cdot\mathbf{M},Z)$ as
\begin{equation}
P(x|\mathbf{a}\cdot\mathbf{M},Z)=P(x|\theta,Z)=\frac{1+xE(\theta)}{2}
\quad,\quad
E(\theta)=E(\mathbf{a}\cdot\mathbf{M},Z)=\sum_{x=\pm1}xP(x|\theta,Z)
,
\label{sg1}
\end{equation}
where $\cos\theta=\mathbf{a}\cdot\mathbf{M}$.
As already mentioned, an essential assumption is that there is no relation between the actual values
of $x_n$ and $x_{n'}$ if $n\not=n'$. With this assumption, repeated application of the product rule yields
\begin{eqnarray}
P(x_1,\ldots,x_N|\mathbf{a}\cdot\mathbf{M},Z)
&=&
\prod_{i=1}^{N}P(x_i|\theta,Z)
.
\label{sg2}
\end{eqnarray}

Although the data set Eq.~(\ref{SG0}) may be expected to change from run to run,
the average Eq.~(\ref{SG1}) should exhibit some kind of robustness with respect to small changes of $\theta$.
Otherwise the average would vary erratically with $\theta$ and
we would discard these ``irreproducible'' results.
Obviously, the expected robustness with respect to small variations
of the conditions under which the experiment is carried out
should be reflected in the expression
for the i-prob to observe the data (within the usual statistical fluctuations).

Let us assume that for a fixed value of $\theta$, an experimental run of $N$ events yields
$n_{x}$ events of the type $\{x\}$ where $n_{+1}+n_{-1}=N$.
The number of different data sets yielding the same values of $n_{+1}$ and $n_{-1}$ is $N!/(n_{+1})!(n_{-1})!$.
The i-prob that events of the type $\{x\}$ occur $n_{x}$ times is given by $\prod_{x\pm1} P(x|\theta,Z)^{n_{x}}$.
Therefore, the i-prob to observe the compound event $\{n_{+1},n_{-1}\}$ is given by
\begin{equation}
P(n_{+1},n_{-1}|\theta,N,Z)=
N!\prod_{x=\pm1} \frac{P(x|\theta,Z)^{n_{x}}}{n_{x}!}
.
\label{robu0}
\end{equation}

If the outcome of the experiment is indeed described by the i-prob
Eq.~(\ref{robu0}) and the experiment is supposed
to yield reproducible, robust results, small changes
of $\theta$ should not have a drastic effect on the outcome.
So let us ask ourselves how the i-prob would change if the experiment is carried
out with $\theta+\epsilon$ ($\epsilon$ small) instead of with $\theta$
by reformulating this question as an hypothesis test.

Let $H_0$ and $H_1$ be the hypothesis that the data $\{n_{+1},n_{-1}\}$ is observed if
the angle between the unit vector $\mathbf{a}$ is $\theta$ and $\theta+\epsilon$, respectively.
The evidence $\mathrm{Ev}$ of hypothesis $H_1$, relative to hypothesis $H_0$, is defined by~\cite{TRIB69,JAYN03}
\begin{equation}
\mathrm{Ev}=\ln\frac{P(n_{+1},n_{-1}|\theta+\epsilon,N,Z)}{P(n_{+1},n_{-1}|\theta,N,Z)}
,
\label{robu1}
\end{equation}
where the logarithm serves to facilitate the algebraic manipulations.
If $H_1$ is more (less) plausible than $H_0$ then $\mathrm{Ev}>0$ ($\mathrm{Ev}<0$).

The absolute value of the evidence, $|\mathrm{Ev}|$ is a measure for the robustness of the description
(the sign of $\mathrm{Ev}$ is arbitrary, hence irrelevant).
The problem of determining the most robust description of the experimental data may now be formulated
as follows: search for the i-prob's $P(n_{+1},n_{-1}|\theta,N,Z)$
which minimize $|\mathrm{Ev}|$ for all possible $\epsilon$ ($\epsilon$ small) and for all possible $\theta$.
The condition ``for all possible $\epsilon$ and  $\theta$'' renders the
minimization problem an instance of a robust optimization problem~\cite{WIKIROBUST}.
Obviously, the robust optimization problem has a trivial solution, namely
$P(n_{+1},n_{-1}|\theta,N,Z)$ independent of $\theta$.
For the case at hand, such $P(n_{+1},n_{-1}|\theta,N,Z)$'s
can only describe experiments for which $\{n_{+1},n_{-1}\}$ show no dependence on $\theta$.
As experiments which produce results that do not change with the conditions
seem fairly pointless, we explicitly exclude solutions
for the i-probs that are constant with respect to the conditions.
It is not difficult to show~\cite{RAED14b} that our concept of a robust experiment implies that the i-prob's which
describe such experiment can be found by minimizing $|\mathrm{Ev}|$, subject to the constraints that
(C1) $\epsilon$ is small but arbitrary, (C2) not all i-prob's are independent of $\theta$, and
(C3) $|\mathrm{Ev}|$ is independent of $\theta$.
The same notion of a robust experiment is the basis for deriving the
Schr\"odinger and Pauli equation and the probabilistic description of the EPRB experiment~\cite{RAED13b,RAED14b,RAED15b}.

Omitting terms of ${\cal O}(\epsilon^3)$, minimizing $|\mathrm{Ev}|$ while taking into account the
constraints (C2) and (C3) amounts to finding the i-prob's $P(x|\theta,Z)$ which minimize~\cite{RAED14b}
\begin{equation}
I_{F}= \sum_{x\pm1}
\frac{1}{P(x|\theta,Z)}
\left(\frac{\partial P(x|\theta,Z)}{\partial\theta}\right)^2
,
\label{robu6}
\end{equation}
subject to the constraint that $\partial P(x|\theta,Z)/\partial\theta\not=0$.
The r.h.s. of Eq.~(\ref{robu6}) is the Fisher information
for the problem at hand. Because of the constraint (C3), it should not depend on $\theta$.
In the course of deriving Eq.~(\ref{robu6}),
our criterion of robustness enforces the intuitively obvious assignment $P(x|\theta,Z)=n_{x}/N$,
which removes the ``subjective'' character of the initial assignment~\cite{RAED14b}.

Using Eq.~(\ref{sg1}), we can rewrite Eq.~(\ref{robu6}) as
\begin{equation}
I_F=\frac{1}{1-E^2(\theta)}
\left(\frac{\partial E(\theta)}{\partial \theta}\right)^2
,
\label{robu6a}
\end{equation}
which is readily integrated to yield
$E(\theta)=\cos\left(\theta\sqrt{I_F}+\phi\right)$
where $\phi$ is an integration constant.
As $E(\theta)$ is a periodic function of $\theta$
we must have $\sqrt{I_F}=K$ where $K$ is an integer and hence
\begin{equation}
E(\theta)=\cos\left(K\theta+\phi\right)
.
\label{robu8}
\end{equation}
Because of constraint (C2)
we exclude the case $K=I_F=0$ from further consideration
because it describes an experiment in which
the frequency distribution of the observed data
does not depend on $\theta$.
Therefore, the physically relevant,
nontrivial solution with minimum Fisher information
corresponds to $K=1$.
Furthermore, as $E(\theta)$ is a function of
$\mathbf{a}\cdot\mathbf{M}=\cos\theta$ only, we must have $\phi=0,\pi$.
\begin{eqnarray}
P(x|\mathbf{a}\cdot\mathbf{M},Z)=
P(x|\theta,Z)=\frac{1\pm x\mathbf{a}\cdot\mathbf{M}}{2}
,
\label{sg5}
\end{eqnarray}
in agreement with the expressions of the quantum theoretical
expression for the probability to deflect the particle
in one of the two distinct directions labeled by $x=\pm1$~\cite{BALL03}.
The $\pm$ sign in Eq.~(\ref{sg5}) reflects the fact that
the mapping between $x=\pm1$ and the two different directions
is only determined up to a sign.

Comparing Eq.~(\ref{sg5}) with the quantum theoretical expression (which is exactly the same)
demonstrates that Born's rule, one of the postulates
of quantum theory, appears as a consequence of LI applied to robust experiments,
a result which not only holds for the SG but also for the EPRB experiment, the Schr\"odinger
and Pauli equation as well~\cite{RAED14b,RAED15b}.
In the LI approach, Eq.~(\ref{sg5}) is not postulated
but follows from the assumption that the (thought) experiment that is being
performed yields the most reproducible results,
revealing the conditions for an experiment to produce data which
is described by quantum theory.
This answers the question ``what is so special about the case in which the separation   procedure
can be carried out and under which conditions can quantum theory be used to describe the statistical experiment?''

\begin{figure}
\includegraphics[width=\hsize]{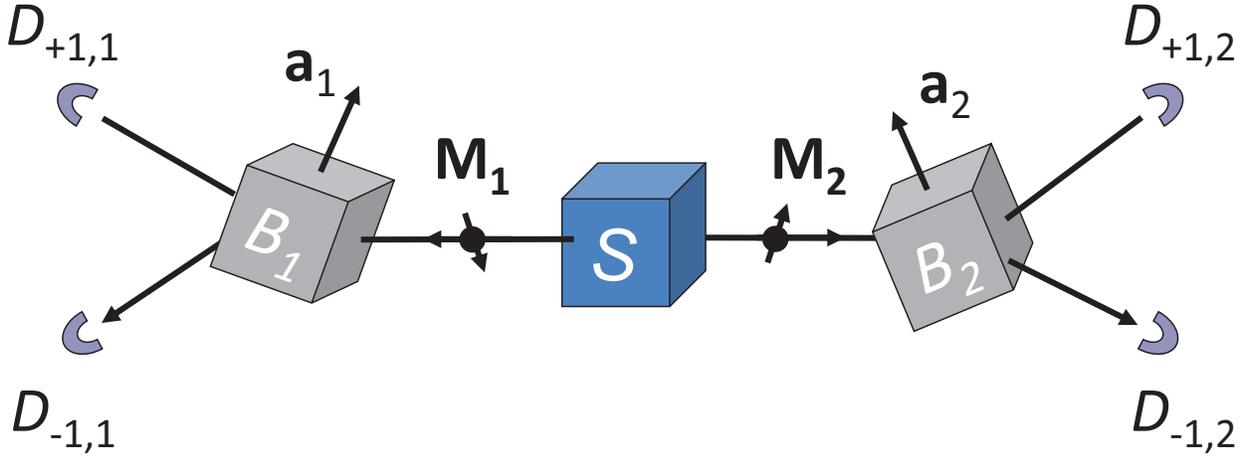}
\caption{(Color online)
Diagram of the EPRB thought experiment.
The source $S$, activated at times labeled by $i=1,2,\ldots,N$,
sends a particle with magnetic moment, represented by the unit vector $\mathbf{M_1}$,
to the Stern-Gerlach magnet $B_1$ and another particle
with magnetic moment, represented by the unit vector $\mathbf{M_2}$, to the Stern-Gerlach magnet $B_2$.
The orientations of the magnets, represented by unit vectors $\mathbf{a_1}$ and $\mathbf{a_2}$,
affect the deflection of the particles by the magnets.
Each particle going to the left (right) is detected with 100\% certainty by either
$D_{+1,1}$ or $D_{-1,1}$ ($D_{+1,2}$ or $D_{-1,2}$).
}
\label{EPRBthought}
\end{figure}

\section{Einstein-Podolsky-Rosen-Bohm thought experiment}\label{sec3}

In this section we repeat the analysis of Section~\ref{sec2}
for Bohm's version of the Einstein-Podolsky-Rosen thought experiment~\cite{EPR35,BOHM51}.
The layout and description of the EPRB thought experiment is given in Fig.~\ref{EPRBthought}.
The result of a run of the experiment for fixed $\mathbf{a}_1$ and $\mathbf{a}_2$ is a data set of pairs
\begin{eqnarray}
{\cal D}&=&\{(x_1,y_1),\ldots,(x_N,y_N)|;x_i=\pm1,y_i=\pm1 ,\;i=1,\ldots,N\}
,
\label{prop0}
\end{eqnarray}
where $N$ is the total number of pairs emitted by the source.
From the data set Eq.~(\ref{prop0}), we compute the numbers of coincidences
\begin{eqnarray}
n_{xy} = \sum_{i=1}^N \delta_{x,x_i}\delta_{y,y_i}\quad,\quad x,y=\pm1
,
\end{eqnarray}
the averages
\begin{eqnarray}
\langle x \rangle = \frac{1}{N}\sum_{i=1}^N x_i=\frac{n_{+1,+1}+n_{+1,-1}-n_{-1,+1}-n_{-1,-1}}{n_{+1,+1}+n_{+1,-1}+n_{-1,+1}+n_{-1,-1}}
\quad,\quad
\label{prop0a}
\langle y \rangle = \frac{1}{N}\sum_{i=1}^N y_i=\frac{n_{+1,+1}+n_{-1,+1}-n_{+1,-1}-n_{-1,-1}}{n_{+1,+1}+n_{+1,-1}+n_{-1,+1}+n_{-1,-1}}
\label{prop0b}
,
\end{eqnarray}
and the correlation
\begin{eqnarray}
\langle xy \rangle &=& \frac{1}{N}\sum_{i=1}^N x_iy_i
=\frac{n_{+1,+1}+n_{-1,-1}-n_{-1,+1}-n_{+1,-1}}{n_{+1,+1}+n_{+1,-1}+n_{-1,+1}+n_{-1,-1}}
.
\label{prop0c}
\end{eqnarray}
As in the case of the SG  experiment, we assume that outcomes with different index $i$ are independent.
In this case, the order in which the pairs appear from the source is irrelevant
and we can compress/represent the content of the data set into/by the four numbers $n_{xy}$ for $x,y=\pm1$.
In other words, the frequencies
\begin{eqnarray}
f(x,y|\mathbf{a}_1,\mathbf{a}_2,\mathbf{M}_1,\mathbf{M}_2,Z)&=&\frac{n_{xy}}{N}
=\frac{1+x\langle x\rangle+y\langle y\rangle + xy\langle xy\rangle}{4}
,
\label{prop0d}
\end{eqnarray}
provide a complete characterization of the data set Eq.~(\ref{prop0})
in terms of the three numbers $\langle x\rangle$, $\langle y\rangle$ and $\langle xy\rangle$.

Instead of continuing with the most general case it is, for pedagogical reasons, expedient
to limit the discussion to the case which is relevant for the EPRB thought experiment.
A first characteristic of the EPRB thought experiment is that
the events $x_i$ and $y_i$ appear random and equally likely.
This implies that $\langle x\rangle=\langle y\rangle=0$.
A second characteristic is that the correlation
$\langle xy\rangle$ does not depends on the orientation of the reference frame.
This implies that $\langle xy\rangle$ must be a function of $\mathbf{a}_1\cdot\mathbf{a}_2$ only.
Therefore, in this special but important case, Eq.~(\ref{prop0d}) takes the form
\begin{eqnarray}
f(x,y|\mathbf{a}_1\cdot\mathbf{a}_2,Z)&=&\frac{n_{xy}}{N}
=\frac{1+ xyE(\mathbf{a}_1\cdot\mathbf{a}_2,Z)}{4}
=\frac{1+ xyE(\theta,Z)}{4}
,
\label{prop1}
\end{eqnarray}
where $\cos\theta=\mathbf{a}_1\cdot\mathbf{a}_2$.
As in the case of the SG experiment, on the basis of Eq.~(\ref{prop1}), there is no way to separate   the
description of the source emitting the particles from the description of the measurement process.
In the subsection that follows, we demonstrate how the quantum theoretical description of the EPRB thought
experiment naturally appears by separating    the description of the data into source and measurement parts.
In essence, we repeat the steps of Section~\ref{sec2a} except that we have four
($x,y=\pm1$) instead of two ($x=\pm1$) possibilities to consider.

\subsection{Separation  procedure}\label{sec3a}

In essence, the following is a straightforward extension of the procedure
outlined in Section~\ref{sec2a}.
We start by writing the observations $xy=(+1,-1,+1,-1)$
and $(f(+1,+1|\theta,Z)$, $f(-1,+1|\theta,Z)$, $f(+1,-1|\theta,Z)$, $f(-1,-1|\theta,Z))$
as $4\times4$ matrices $X$ and $F$ with elements
$X([x,y],[x',y'])=x\delta_{x,x'}\delta_{y,y'}$ and
$Y([x,y],[x',y'])=y\delta_{x,x'}\delta_{y,y'}$ and
$F([x,y],[x',y'])=f(xy|\mathbf{a},\mathbf{M},Z)\delta_{x,x'}\delta_{y,y'}$, respectively.
Here we use the notation $[x,y]=(1-x)/2+(1-y)$ to indicate that the pairs $(x,y)$ and
$(x',y')$ specify the row, respectively the column index (running from 0 to 3)
of the matrices $X$ and $F$.
We require that
\begin{eqnarray}
\mathbf{Tr} \widehat\rho =1
\quad,\quad
\mathbf{Tr} \widehat\rho \widehat X=\langle x\rangle
\quad,\quad
\mathbf{Tr} \widehat\rho \widehat Y=\langle y\rangle
\quad,\quad
\mathbf{Tr} \widehat\rho \widehat X \widehat Y=\langle xy\rangle
.
\label{sec3a0}
\end{eqnarray}
where instead of the $2\times2$ matrices in the SG case, we now look for $4\times4$
matrices $\widehat\rho$, $\widehat X$, and  $\widehat Y$
which satisfy Eq.~(\ref{sec3a0}) but allow for the desired separation.
From linear algebra we know that any hermitian $4\times4$ matrix
can be written as a linear combination of sixteen hermitian $4\times4$ matrices.
These sixteen matrices may be chosen such that they are orthonormal
with respect to the inner product defined by $(A,B)=\mathbf{Tr\;} A^\dagger B$.
Using the direct product of the Pauli-spin matrices
$\bm{\sigma}_j=(\sigma^x_j,\sigma^y_j,\sigma^z_j)$ for $j=1,2$
and the unit matrix $\openone$ as the orthonormal basis set for the vector space of
$4\times4$ matrices, we may write (without loss of generality)
\begin{eqnarray}
\widehat\rho&=&\rho_0\openone+\bm\rho_1\cdot\bm{\sigma}_1\otimes\openone_2+\openone_1\otimes\bm\rho_2\cdot\bm{\sigma}_2
+\bm{\sigma}_1\cdot\bm\rho_{12}\cdot\bm{\sigma}_2
.
\label{P2D1a}
\end{eqnarray}
where the number $\rho_0$, the vectors $\bm\rho_j$, and the matrix $\bm\rho_{12}$ are all real-valued.
As each of the two sides of the EPRB experiment contains a SG magnet,
consistency with the separated   description of the SG experiment demands that we choose
\begin{equation}
\widehat X=\mathbf{a}_1\cdot\bm{\sigma}_1\otimes\openone_2
\quad,\quad \widehat Y=\openone_1\otimes\mathbf{a}_2\cdot\bm{\sigma}_2
.
\label{P2D10}
\end{equation}
We find the explicit expression of $\widehat\rho$ by requiring that Eq.~(\ref{sec3a0}) holds.
Focussing on the case of the EPRB experiment
for which $\langle x\rangle=\langle y\rangle=0$ and
$\langle xy\rangle=-\mathbf{a}_1\cdot\mathbf{a}_2$,
it readily follows that $\rho_0=1/4$, $\bm\rho_1=\bm\rho_2=0$
and that $\widehat\rho$ takes the form
\begin{eqnarray}
\widehat\rho&=&\frac{\openone-\sigma_1\cdot\sigma_2}{4}
.
\label{P2D12}
\end{eqnarray}
It is not difficult to verify that ${\widehat\rho}^2={\widehat\rho}$, hence, in quantum parlance,
Eq.~(\ref{P2D12}) is the density matrix of a pure quantum state~\cite{BALL03}.
Computing the matrix elements of ${\widehat\rho}$ in the
spin-up, spin-down basis of both spins we find
\begin{eqnarray}
\widehat\rho&=&
\left(\frac{|\uparrow\downarrow\rangle-|\downarrow\uparrow\rangle}{\sqrt{2}}\right)
\left(\frac{\langle\uparrow\downarrow|-\langle\downarrow\uparrow|}{\sqrt{2}}\right)
,
\label{P2D12a}
\end{eqnarray}
and
\begin{eqnarray}
\langle xy\rangle&=&\mathbf{Tr\;} \widehat\rho \widehat X \widehat Y
=\mathbf{Tr\;} \widehat\rho \mathbf{a}_1\cdot\bm{\sigma}_1\mathbf{a}_2\cdot\bm{\sigma}_2
=
-\mathbf{a}_1\cdot\mathbf{a}_2
,
\label{P2D12b}
\end{eqnarray}
which we recognize as the quantum theoretical description of two spin-1/2 objects in the singlet state.

In other words, we have shown that rewriting the data
gathered in an ideal EPRB thought experiment in a manner
that allows for the envisaged separation   naturally leads,
without invoking postulates of quantum theory,
to the quantum theoretical problem of two $S=1/2$ spins in the singlet state.

As in the case of the ideal SG experiment, the separated   representation Eq.~(\ref{P2D10}) and Eq.~(\ref{P2D12})
puts a severe restriction on the kind of data that we can describe, again
provoking to question ``what is so special about the case in which the separation   procedure
can be carried out?''
In the next subsection, we answer this question
in terms of a LI treatment of robust EPRB experiments~\cite{RAED14b}.

\subsection{Logical-inference treatment}\label{sec3b}
A detailed account of the LI treatment of the EPRB thought experiment can be found
in {Ref.~\onlinecite{RAED14b}}.
Moreover, as the LI treatment of the EPRB thought experiment is,
in essence, the same as the one of the idealized SG experiment given in Section~\ref{sec2b},
we limit the presentation to a discussion of the assumptions and the results.

\begin{enumerate}[1.]
\item
The i-prob to observe a pair $\{x,y\}$
is denoted by $P(x,y|\mathbf{a}_1,\mathbf{a}_2,Z)$
where $Z$ represents all the conditions
under which the experiment is performed, with exception of
the directions $\mathbf{a}_1$ and $\mathbf{a}_2$ of the magnets $B_1$ and $B_2$, respectively.
It is assumed that the conditions represented by $Z$ are fixed and identical
for all experiments.
Note that we have omitted the (in)dependence on $M_1$ and $M_2$ because in the case
at hand, it is redundant~\cite{RAED14b}.
\item
For simplicity,
it is assumed that there is no relation between the actual values
of the pairs $\{x_n,y_n\}$ and $\{x_{n'},y_{n'}\}$ if $n\not=n'$,
meaning that each repetition
of the experiment represents an identical event of which
the outcome is logically independent of any other such event.
Invoking the product rule, the logical consequence of this assumption is that
\begin{eqnarray}
P(x_1,y_1,\ldots,x_N,y_N|\mathbf{a}_1,\mathbf{a}_2,Z)&=&\prod_{i=1}^{N}P(x_i,y_i|\mathbf{a}_1,\mathbf{a}_2,Z)
,
\label{prop2}
\end{eqnarray}
meaning that the i-prob
$P(x_1,y_1,\ldots,x_N,y_N|\mathbf{a}_1,\mathbf{a}_2,Z)$
to observe the compound event $\left\{\{x_1,y_1\},\ldots,\{x_N,y_N\}\right\}$
is completely determined by the i-prob $P(x,y|\mathbf{a}_1,\mathbf{a}_2,Z)$ to observe the pair $\{x,y\}$.
\item
It is assumed that the i-prob $P(x,y|\mathbf{a}_1,\mathbf{a}_2,Z)$
to observe a pair $\{x,y\}$ does not change if we apply
the same rotation to both magnets $B_1$ and $B_2$.
Expressing this invariance with respect to rotations
of the coordinate system (Euclidean space and Cartesian coordinates are used throughout this paper)
in terms of i-probs requires that
$P(x,y|\mathbf{a}_1,\mathbf{a}_2,Z)=P(x,y|{\cal R}\mathbf{a}_1,{\cal R}\mathbf{a}_2,Z)$
where ${\cal R}$ denotes an arbitrary rotation in three-dimensional space which is
applied to both magnets $B_1$ and $B_2$.
As a function of the vectors $\mathbf{a}_1$ and $\mathbf{a}_2$,
the functional equation
$P(x,y|\mathbf{a}_1,\mathbf{a}_2,Z)=P(x,y|{\cal R}\mathbf{a}_1,{\cal R}\mathbf{a}_2,Z)$
can only be satisfied for all $\mathbf{a}_1$, $\mathbf{a}_2$ and
rotations ${\cal R}$ if $P(x,y|\mathbf{a}_1,\mathbf{a}_2,Z)$
is a function of the inner product $\mathbf{a}_1\cdot\mathbf{a}_2$ only.
Therefore, we must have
\begin{equation}
P(x,y|\mathbf{a}_1,\mathbf{a}_2,Z)=P(x,y|\mathbf{a}_1\cdot\mathbf{a}_2,Z)=P(x,y|\theta,Z)
,
\label{prop3}
\end{equation}
where $\theta=\arccos(\mathbf{a}_1\cdot\mathbf{a}_2)$ denotes the angle between the unit vectors
$\mathbf{a}_1$ and $\mathbf{a}_2$.
For any integer value of $K$, $\theta+2\pi K$
represents the same physical arrangement of the magnets $M_1$ and $M_2$.
\item
From the LI rules, it follows that the i-prob to observe $x$, irrespective of the observed value of $y$ is given by
\begin{equation}
P(x|\mathbf{a}_1,\mathbf{a}_2,Z)=\sum_{y=\pm1}P(x,y|\mathbf{a}_1,\mathbf{a}_2,Z)
.
\label{prop3a}
\end{equation}
The assumption that observing $x=+1$ is as likely as
observing $x=-1$, independent of the observed value of $y$,
implies that we must have $P(x=+1|\mathbf{a}_1,\mathbf{a}_2,Z)=P(x=-1|\mathbf{a}_1,\mathbf{a}_2,Z)$ which,
in view of the fact that $P(x=+1|\mathbf{a}_1,\mathbf{a}_2,Z)+P(x=-1|\mathbf{a}_1,\mathbf{a}_2,Z)=1$
implies that $P(x=+1|\mathbf{a}_1,\mathbf{a}_2,Z)=P(x=-1|\mathbf{a}_1,\mathbf{a}_2,Z)=1/2$.
Applying the same reasoning to the assumption that, independent of the observed values of $x$,
observing $y=+1$ is as likely as observing $y=-1$
yields $P(y|\mathbf{a}_1,\mathbf{a}_2,Z)=P(x=+1,y|\mathbf{a}_1,\mathbf{a}_2,Z)+P(x=-1,y|\mathbf{a}_1,\mathbf{a}_2,Z)=1/2$.
\end{enumerate}

\noindent
Note that we did not assign any prior i-prob nor did we make any reference to concepts such as the singlet-state.
Although the symmetry properties which have been assumed are reminiscent of those of the singlet state,
this is deceptive: without altering the assumptions that are expressed in (3) and (4),
the LI approach yields the correlations for the triplet states as well~\cite{RAED14b}.

Adopting the same reasoning as in Section~\ref{sec2b},
it follows directly from assumptions (1--4)  that the i-prob to observe a pair $\{x,y\}$
takes the form~\cite{RAED14b}
\begin{equation}
P(x,y|\theta,Z)=\frac{1+xyE_{12}(\theta)}{4}
,
\label{prop5}
\end{equation}
where
$E_{12}(\theta)=E_{12}(\mathbf{a}_1,\mathbf{a}_2,Z)$ is a periodic function of $\theta$.
Minimization of the corresponding expression of $|\mathrm{Ev}|$ while taking into account the
constraints (C2) and (C3) (see Section~\ref{sec2b})
is tantamount to finding the i-prob's $P(x,y|\theta,Z)$ which minimize~\cite{RAED14b}
\begin{equation}
I_{F}= \sum_{x,y=\pm1}
\frac{1}{P(x,y|\theta,Z)}
\left(\frac{\partial P(x,y|\theta,Z)}{\partial\theta}\right)^2
,
\label{epr6}
\end{equation}
subject to the constraint that $\partial P(x,y|\theta,Z)/\partial\theta\not=0$
for some pairs $(x,y)$.
The r.h.s. of Eq.~(\ref{epr6}) is the Fisher information
for the problem at hand and because of constraint (C3), does not depend on $\theta$.
Using Eq.~(\ref{prop5}), we can rewrite Eq.~(\ref{epr6}) as
\begin{equation}
I_F=\frac{1}{1-E_{12}^2(\theta)}
\left(\frac{\partial E_{12}(\theta)}{\partial \theta}\right)^2
,
\label{epr6a}
\end{equation}
which is readily integrated to yield
$E_{12}(\theta)=\cos\left(\theta\sqrt{I_F}+\phi\right)$
where $\phi$ is an integration constant.
As $E_{12}(\theta)$ is a periodic function of $\theta$
we must have $\sqrt{I_F}=K$ where $K$ is an integer and hence
\begin{equation}
E_{12}(\theta)=\cos\left(K\theta+\phi\right)
.
\label{epr8}
\end{equation}
Because of constraint (C2)
we exclude the case $K=I_F=0$ from further consideration
because it describes an experiment in which
the frequency distribution of the observed data
does not depend on $\theta$.
Therefore, the physically relevant,
nontrivial solution with minimum Fisher information
corresponds to $K=1$.
Furthermore, as $E_{12}(\theta)$ is a function of
$\mathbf{a}_1\cdot\mathbf{a}_2=\cos\theta$ only,
we must have $\phi=0,\pi$, reflecting an ambiguity
in the definition of the direction
of $B_1$ relative to the direction of $B_2$.

Choosing the solution with $\phi=\pi$, the two-particle i-prob reads
\begin{eqnarray}
P(x,y|\mathbf{a}_1,\mathbf{a}_2,Z)=\frac{1+xy\mathbf{a}_1\cdot\mathbf{a}_2}{4}
\quad,\quad
\sum_{x,y=\pm1} xyP(x,y|\mathbf{a}_1,\mathbf{a}_2,Z)=\langle xy\rangle =
-\mathbf{a}_1\cdot\mathbf{a}_2
,
\label{P2D9}
\label{P2D9a}
\end{eqnarray}
in agreement with the expression for the correlation of two $S=1/2$
particles in the singlet state~\cite{BOHM51,BALL03}.
We may therefore conclude that without making reference to concepts of quantum theory,
the LI treatment of the robust EPPB experiment directly yields the probabilistic
description that we know from quantum theory.
This demonstrates that the data produced by EPRB experiments can be described,
in a more general manner, without invoking the notion of quantum entanglement.

\subsection{Discussion}\label{sec3c}

In Section~\ref{sec3a} we apply the separating   procedure to the EPRB thought experiment and demonstrate
how the quantum theoretical description in terms of the singlet state emerges from a simple
rearrangement of the representation of the data.

Section~\ref{sec3b} shows that the application of the criterion of robust, reproducible experiments~\cite{RAED14b}
to the EPRB thought experiment depicted in Fig.~\ref{EPRBthought}
amounts to minimizing the Fisher information Eq.~(\ref{epr6a}) for this specific problem.
The result of this calculation is Eq.~(\ref{P2D9}),
the probability distribution that is characteristic for two spin-1/2 objects in the singlet state.
Needless to say, the derivation that led to Eq.~(\ref{P2D9}) did not require invoking concepts of quantum theory.
Only rational reasoning strictly complying with the rules of LI and some elementary facts about the experiment were used.

It is most remarkable that the quantum theoretical description of a system in the singlet state emerges by simply
requiring that (i) everything which is known about the source
is uncertain, except that it emits two signals, (ii) the magnets
$B_1$ and $B_2$ transform the received signal
into two-valued signals, and that
(iii) the i-prob describing the frequencies of the pair of events $(x,y)$ does not depend on the orientation of the reference frame~\cite{RAED14b}.
We emphasize that Sections 3.1 and 3.2 address different ways of representing frequencies of
two-valued data obtained in robust experiments and have no bearing on the issue non-locality as such.

In spite of the widely spread claims that real EPRB experiments
have proven quantum theory correct, it is a fact that none of the three different
experiments for which data has been made available~\cite{WEIH00,SHIH11,VIST12}
survives the confrontation with the 5-standard-deviation-criterion
hypothesis test that the data complies with the quantum theoretical description Eq.~(\ref{P2D9})~\cite{RAED13a}.
It seems that for the time being, only computer experiments are able to
generate data that are not in conflict with the quantum theoretical description
of the EPRB thought experiment~\cite{RAED13a}.

\section{Relation to previous work}\label{RELA}

\subsection{Subject-object separation}

The general issue of separating subject and object or more specifically, the limitations on the separation of
the phenomenon under scrutiny from the process of gathering data about it
have played an important role in the development of quantum physics~\cite{BOHR79,BOHR99,PLOT10a,PLOT13}.
In this context, the typical signatures of quantum physics appear as result of the impossibility
to gather data about the (atomic) objects without significantly disturbing their behavior.
While this viewpoint is implicit in the LI derivation of equations such as the Schr\"odinger equation~\cite{RAED13b,RAED14b,RAED15b},
in our view, the separation  procedure discussed in the present paper does not really address this issue.
The separation  procedure explores alternatives for
representing data sets and selects the quantum theoretical description as the representation in which
the various parts of the experiment that may affect the data production process
appear as independent entities, in sharp contrast with the original
representation of the data set in terms of frequencies.
On the other hand, there is a relation between the notion of ``complementarity'' advocated by N. Bohr~\cite{BOHR79,BOHR99,PLOT10a,PLOT13}
and the separation  procedure discussed in the present paper.

For simplicity, we discuss this issue using the SG experiment as an example but without modification, the arguments carry over to the EPRB experiment as well.
Suppose that we want to describe the outcomes of two SG experiments with two different settings $\mathbf{a}\not=\mathbf{b}$.
Obviously, in practice we cannot perform the experiment with $\mathbf{a}$ and $\mathbf{b}$ at the same time: the two conditions are mutually exclusive.
On the level of the description in terms of the frequency Eq.~(\ref{INT1}),
we require the specification of two numbers $E(\mathbf{a},\mathbf{M},Z)$ and $E(\mathbf{b},\mathbf{M},Z)$.
In general, there is no principle, in fact no reason at all, that would enforce a relation between these two numbers:
a full characterization of the experiment in terms of classical concepts (the directions $\mathbf{a}$ and $\mathbf{M}$)
would require data of $E(\mathbf{a},\mathbf{M},Z)$ for all $\mathbf{a}$ (for simplicity we assume that $\mathbf{M}$ is fixed).
But once we require that the separation  procedure can be carried out, the matrix calculus that
is characteristic of quantum theory emerges and the resulting formalism in terms of non-commuting matrices
allows for one common description of {\bf all} possible mutually exclusive experiments.
Apparently, the conditions for which Bohr's statement\cite{BOHR79} ``an adequate tool for a complementary
way of description is offered precisely by the quantum-mechanical formalism which represents a purely symbolic
scheme permitting only predictions, on lines of the correspondence
principle, as to results obtainable under conditions specified by means of classical concepts''
holds correspond to those for which the separation  procedure can be carried out.

\subsection{Gleason-Bush theorem}

The Gleason-Bush theorem~\cite{GLEA57,BUSC03} assumes that there is a (finite) Hilbert space,
with ``events'' being defined to be closed subspaces of this Hilbert space.
One then considers the algebra of these ``events'' and defines a map (a kind of  ``probability'') on this algebra, mapping events to real numbers.
The theorem then says that this map must have the structure
$\langle \mathbf{v}| \rho |  \mathbf{v} \rangle$ where $ \mathbf{v}$
represents one of the ``events'' and $\rho$ is a density matrix.
The key point is that the algebra dictates the structure of the map (i.e. Born's rule).
Based on these considerations, Pitowsky suggests that the Hilbert space formalism of
quantum mechanics is a new theory of probability~\cite{PITO06} and continues to state that
``The quantum structure is in this sense much more constrained than the classical formalism.
The structure of the phase space of a classical system does not greatly restrict the type of probability measures
that can be defined on it. The probability measures which are actually used
in classical statistical mechanics are introduced mostly by fiat or, in any case,
are very hard to justify.''
This line of thinking is orthogonal to the one adopted in the present paper
in which events are not restricted to be ``closed subspaces of some Hilbert space''.
By way of concrete examples, the present paper shows that the Hilbert space formalism emerges from the desire to separate
the description into various parts and that the quantum theoretical description is indeed a constrained form of probability theory.

\section{Summary}\label{CONC}\label{DISC}

In Section~\ref{sec2a}, the quantum theoretical description of the idealized SG experiment
is shown to derive by separating   the representation of the observed data
into a description of the object under study and a description of the measurement device.
Section~\ref{sec2b} shows that the class of experimental outcomes which can
be described by quantum theory are special in the sense that they
are the results of a robust Stern-Gerlach experiment~\cite{RAED14b}.

Section~\ref{sec3a} demonstrates that the quantum theoretical description of the
Einstein-Podolsky-Rosen-Bohm thought experiment naturally emerges by
separating   the representation of the observed data
into a description of the object under study and a description of the measurement device.
As the logical-inference derivation of Section~\ref{sec3b} yields the same expressions,
it also follows that the class of experiments which quantum theory can describe
is smaller than the one which allows a description in terms of conditional probabilities.

The answer to the question ``what is so special about the case in which the separation   procedure
can be carried out and under which conditions can quantum theory be used to describe the statistical experiment?''
is the same for both the Stern-Gerlach and Einstein-Podolsky-Rosen-Bohm thought experiment, namely
that the experimental data which quantum theory can describe are special in the sense that they
are the result of a robust experiment~\cite{RAED14b}.
Once we accept as a principle, the idea of separation in the sense explained in Section~\ref{sec2a},
the representation of the data in terms of quantum theoretical constructs is
completely determined by the rules of linear algebra, i.e. by mathematics alone.

Although logical-inference applied to a different class of robust experiments as those discussed in the present paper
lead to the Schr\"odinger and Pauli equation~\cite{RAED13b,RAED14b,RAED15b},
in the current formulation there is nothing to separate: there is only a data set
of clicks of detectors and the no control over the unknown position of the particle~\cite{RAED14b,RAED15b}.
Extending the application of the  separation procedure
to these more complicated cases is left for future research.

\section*{Acknowledgement}
We are grateful to Karl Hess and Arkady Plotnitsky for commenting on an early version of this paper.
We thank Koen De Raedt for many stimulating discussions.
MIK and HCD acknowledges financial support by European Research
Council, project 338957 FEMTO/NANO.

\bibliographystyle{spiebib}   
\bibliography{/d/papers/all13}

\end{document}